\documentclass[conference]{IEEEtran}
\usepackage[T1]{fontenc}
\usepackage{cite}
\usepackage{amsmath,amssymb}
\usepackage{algorithm}
\usepackage[noend]{algpseudocode}
\usepackage{graphicx}
\usepackage{booktabs}
\usepackage{multirow}
\usepackage{threeparttable}
\usepackage{tabularx}
\usepackage{array}
\usepackage{balance}
\usepackage{url}
\usepackage{microtype}
\usepackage[dvipsnames]{xcolor}
\usepackage{tikz}
\usepackage{pgfplots}
\usepackage{pgfplotstable}
\usetikzlibrary{arrows.meta,positioning,calc,fit,backgrounds,patterns}
\usepgfplotslibrary{groupplots,statistics}
\pgfplotsset{compat=1.18}

\renewcommand{\texttt}[1]{{\ttfamily\bfseries #1}}
\newcommand{\code}[1]{\texttt{#1}}
\newcommand{\vect}[1]{\boldsymbol{#1}}

\newcommand{\cB}{\mathcal{B}}
\newcommand{\cC}{\mathcal{C}}
\newcommand{\cF}{\mathcal{F}}
\newcommand{\cG}{\mathcal{G}}
\newcommand{\cL}{\mathcal{L}}
\newcommand{\cT}{\mathcal{T}}
\newcommand{\firstK}{\operatorname{first}_K}
\newcommand{\figref}[1]{Fig.~\ref{#1}}
\newcommand{\Figref}[1]{Figure~\ref{#1}}

\newcommand{\Tabref}[1]{Table~\ref{#1}}

\newcommand{\Secref}[1]{Section~\ref{#1}}
\newcommand{\eqnref}[1]{(\ref{#1})}
\newcommand{\Algref}[1]{Algorithm~\ref{#1}}

\begin{document}

\title{Feasibility-Aware Security-Constrained Unit Commitment via Hybrid Soft Actor-Critic with Quantum-Sampled Features}

\author{\IEEEauthorblockN{George Dimas, Amin Masoumi, and Mert Korkali}
\IEEEauthorblockA{\textit{Department of Electrical Engineering and Computer Science}\\
\textit{University of Missouri}\\
Columbia, MO 65211 USA\\
E-mail: {\{gadtbg, am4n5, korkalim\}@missouri.edu}}
}

\maketitle
\bstctlcite{IEEEexample:BSTcontrol}

\begin{abstract}
Security-constrained unit commitment (SCUC) couples binary commitment, economic dispatch, reserves, and network security over a multiperiod horizon, which makes an exact solution expensive at realistic system sizes. This paper proposes a three-layer hybrid framework in which a Bernoulli hybrid soft actor-critic (HSAC) policy proposes hourly commitments, a quantum-sampled auxiliary channel augments the state, and a native SCUC mixed-integer linear program recovers dispatch and security variables after only a limited subset of commitment binaries is enforced. The method is therefore solver-compatible rather than an end-to-end replacement for exact optimization. We formalize the SCUC-to-reinforcement-learning interface, derive the temporal coverage induced by the fixed cap, and conduct representative experiments on the 14-, 57-, and 118-bus cases. The results show stable, low-cost recovery in the 14-bus case; a very low screen-rejection rate in the 57-bus case, consistent with learned feasibility generalization under fixed intertemporal SCUC constraints; and a clear coverage bottleneck in the 118-bus case once the enforcement cap no longer spans a complete commitment period. The 118-bus case runtime traces nevertheless remain tightly clustered for accepted episodes, indicating that the policy still captures a repeatable recovery pattern across most episodes. The study, therefore, identifies the dominant limitation of the current implementation as the amount of useful commitment information that reaches the recovery model under an exploratory Bernoulli actor and a small enforcement cap, and shows how that limitation governs scalability.
\end{abstract}


\section{Introduction}
\label{sec:intro}
Security-constrained unit commitment (SCUC) is a central optimization problem in power-system operations and electricity markets because it must coordinate binary generator commitments, continuous dispatch, reserves, and transmission-security constraints over a rolling horizon. Even when the network is linearized with a direct-current (DC) approximation, the resulting mixed-integer linear program (MILP) remains computationally demanding at realistic system scales \cite{chen2023scuc,yang2021ml}. The challenge becomes sharper when operators require repeated solves under renewable uncertainty, tighter reserve margins, and short market-clearing deadlines.
Recent research has explored learning-assisted SCUC through warm starts, branching guidance, integer-variable reduction, and feasibility-oriented screening \cite{xavier2021learning,pineda2022low,dai2025reduction,venkatesh2025feasibility,wang2026structure,xiong2025successive}. Reinforcement learning (RL) has also been applied to unit commitment (UC) through actor-critic policies, tree search, adaptive horizons, optimization-assisted policy learning, and uncertainty-aware dispatch coupling \cite{demars2021applying,demars2022astar,qin2023ensemble,sayed2024acuc,xu2024wind,yan2024lookahead,liang2024expert,lu2026graph}. In parallel, quantum and hybrid quantum-classical studies have examined the quantum approximate optimization algorithm (QAOA), annealing, surrogate decomposition, exact quantum search, and hybrid recovery schemes for UC \cite{koretsky2021qaoa,feng2023surrogate,zheng2024fastquantum,wei2026qrl,hong2025qa,salgado2024hybrid,liu2025exact,aboumrad2025hybrid,gao2025dqgbd,hasanzadeh2025d2uc,hasanzadeh2026survey,muller2026qa,barrass2025leveraging}. These directions are valuable, but two practical gaps remain.
First, end-to-end RL policies often need a repair layer because exact UC feasibility is difficult to enforce directly in policy space, especially when intertemporal and network constraints are active \cite{pineda2022low,sayed2024acuc}. Second, many quantum UC studies rely on quadratic unconstrained binary optimization (QUBO) reformulations or surrogate decompositions whose scalability is still limited by near-term hardware and embedding constraints \cite{koretsky2021qaoa,hong2025qa,aboumrad2025hybrid,hasanzadeh2026survey}. These observations motivate a more conservative integration strategy, namely to retain the native SCUC solver and learn only a compact subset of commitment decisions that meaningfully shrinks the binary search space.
This paper adopts exactly that strategy through a \textit{three-layer hybrid framework}.\footnote{The three layers are policy generation, quantum feature sampling, and classical SCUC recovery.} The proposed method couples a Bernoulli hybrid soft actor-critic (HSAC) policy, a quantum-sampled state-augmentation channel, a capacity screen, and a warm-started SCUC recovery model that enforces only a limited subset of policy-proposed commitment binaries. In the present implementation, commitment tuples are appended in time order, and only the first $K$ tuples are imposed as equality constraints in the recovery model. The main contributions are threefold. First, we formulate a solver-compatible RL-to-SCUC interface in which policy outputs restrict only part of the binary space, while the original SCUC model retains responsibility for dispatch, reserves, and security feasibility. Second, we tailor soft actor-critic (SAC) to multi-binary UC actions for entropy-regularized learning with a mixed-integer recovery layer. Third, we show that the current chronological enforcement rule induces a measurable coverage ratio that strongly influences scalability. 

The remainder of this paper is organized as follows. \Secref{sec:formulation} formulates SCUC and the sequential decision mapping. \Secref{sec:method} describes the proposed HSAC-SCUC pipeline and algorithm. \Secref{sec:setup} summarizes the implementation and experimental setup. \Secref{sec:results} presents the results and discussion. \Secref{sec:conclusion} concludes the paper.

\section{SCUC Formulation and Sequential Decision Mapping}
\label{sec:formulation}
\subsection{SCUC Model}
Let $\cG$, $\cB$, $\cL$, $\cC$, and $\cT=\{1,\ldots,T\}$ denote the sets of thermal units, buses, transmission lines, contingencies, and time periods. We use $N_g=|\cG|$, $N_b=|\cB|$, and $N_l=|\cL|$. For each Bus $b\in\cB$, let $\cG_b\subseteq\cG$ be the set of units connected to $b$, and let $\delta(b)\subseteq\cL$ denote the incident lines. For each Line $l\in\cL$, let $i(l)$ and $j(l)$ denote its sending- and receiving-end buses. For each Generator $g\in\cG$ and Period $t\in\cT$, the binary variables $u_{g,t}$, $v_{g,t}$, and $w_{g,t}$ denote the on/off, startup, and shutdown decisions, respectively. The continuous variables $p_{g,t}$, $r_{g,t}$, $f_{l,t}$, $\theta_{b,t}$, and $\ell_{b,t}^{\mathrm{ls}}$ denote active-power dispatch, reserve, line flow, phase angle, and involuntary load shedding, respectively. The demand at Bus $b$ and Time $t$ is $d_{b,t}$; the total system demand is $D_t=\sum_{b\in\cB} d_{b,t}$; the reserve requirement is $R_t^{\mathrm{req}}$; and $C_g(\cdot)$ denotes the production-cost function of Unit $g$, i.e., a convex or piecewise-linear approximation of dispatch cost.
The SCUC objective minimizes no-load, startup, shutdown, dispatch, and load-shedding costs, viz.,
\begin{align}
\min\; Z &= \sum_{t\in\cT}\sum_{g\in\cG} \Big(c_g^{\mathrm{nl}}u_{g,t}+c_g^{\mathrm{su}}v_{g,t}+c_g^{\mathrm{sd}}w_{g,t}+C_g(p_{g,t})\Big) \nonumber\\
&\quad + \sum_{t\in\cT}\sum_{b\in\cB} c^{\mathrm{ls}}\ell_{b,t}^{\mathrm{ls}}.
\label{eq:obj}
\end{align}
A representative SCUC model used throughout the paper is
\begin{subequations}\label{eq:scuc}
\begin{align}
u_{g,t}-u_{g,t-1} &= v_{g,t}-w_{g,t}, && \forall g,t, \label{eq:logic}\\
\sum_{\tau=t}^{t+UT_g-1} u_{g,\tau} &\ge UT_g v_{g,t}, && \forall g,t, \label{eq:uptime}\\
\sum_{\tau=t}^{t+DT_g-1} (1-u_{g,\tau}) &\ge DT_g w_{g,t}, && \forall g,t, \label{eq:dntime}\\
\underline P_g u_{g,t} \le p_{g,t} &\le \overline P_g u_{g,t}, && \forall g,t, \label{eq:pbound}\\
-RD_g \le p_{g,t}-p_{g,t-1} &\le RU_g, && \forall g,t, \label{eq:ramp}\\
\sum_{g\in\cG_b} p_{g,t} - d_{b,t} + \ell_{b,t}^{\mathrm{ls}} &= \sum_{l\in\delta(b)} f_{l,t}, && \forall b,t, \label{eq:balance}\\
f_{l,t} &= B_l\big(\theta_{i(l),t}-\theta_{j(l),t}\big), && \forall l,t, \label{eq:dcflow}\\
|f_{l,t}| &\le F_{l}^{\max}, && \forall l,t, \label{eq:flowlimit}\\
p_{g,t}+r_{g,t} &\le \overline P_g u_{g,t}, && \forall g,t, \label{eq:headroom}\\
\sum_{g\in\cG} r_{g,t} &\ge R_t^{\mathrm{req}}, && \forall t, \label{eq:reserve}\\
|f_{l,t}^{(c)}| &\le F_{l,c}^{\max}, && \forall l,c,t. \label{eq:security}
\end{align}
\end{subequations}
Equation \eqnref{eq:logic} enforces commitment transitions; \eqnref{eq:uptime}--\eqnref{eq:dntime} impose minimum up- and downtimes; \eqnref{eq:pbound}--\eqnref{eq:headroom} define dispatch, ramping, and reserve headroom limits; and \eqnref{eq:balance}--\eqnref{eq:security} enforce network balance, DC line limits, and post-contingency security. Here, $UT_g$ and $DT_g$ are minimum up- and downtimes; $RU_g$ and $RD_g$ are ramp-up and ramp-down limits; $B_l$ is the line susceptance; and $F_l^{\max}$ and $F_{l,c}^{\max}$ are the base-case and post-contingency flow limits, respectively. As in standard market-grade SCUC formulations, the recovery layer keeps these physical constraints intact \cite{chen2023scuc,xavier2024ucjl}.

\subsection{Sequential Mapping and Reward}
We map partial commitment enforcement to a sequential decision process. At stage $t$, the agent observes
\begin{align}
\vect{s}_t=[\sin(2\pi t/T),\;\cos(2\pi t/T),\;\vect{u}_{t-1},\;\vect{z}_t],
\label{eq:state}
\end{align}
where the sinusoidal pair encodes time of day; $\vect{u}_{t-1}\in\{0,1\}^{N_g}$ is the previous commitment vector; and $\vect{z}_t\in[0,1]^{N_g}$ is a quantum-sampled auxiliary feature vector. The action is a multi-binary commitment proposal $\vect{a}_t\in\{0,1\}^{N_g}$ that is subsequently screened and partially enforced before the mixed-integer recovery solution.
The per-stage reward combines normalized recovery cost, switching, time, and infeasibility penalties with potential-based shaping, viz., 
\begin{align}
r_t^{\mathrm{RL}} &= -\Big(\lambda_c\tfrac{J_t}{\bar J}+\lambda_{\mathrm{sw}}\psi_t^{\mathrm{sw}}+\lambda_{\mathrm{time}}\psi_t^{\mathrm{time}}+\lambda_{\mathrm{inf}}\psi_t^{\mathrm{inf}}\Big) \nonumber\\
&\quad + \lambda_{\Phi}\big(\gamma\Phi_{t+1}-\Phi_t\big),
\label{eq:reward}
\end{align}
with
\begin{align}
\Phi_t = -\kappa\,\widehat C^{\mathrm{ED}}\big(\widehat D_{t+1}^{\mathrm{net}},\vect{u}_t\big)/\bar J.
\label{eq:potential}
\end{align}
Here, $J_t$ is the recovered SCUC objective at Stage $t$; $\bar J$ is a cost normalizer; $\psi_t^{\mathrm{sw}}=N_g^{-1}\lVert\vect{a}_t-\vect{u}_{t-1}\rVert_1$ measures commitment switching; $\xi_t^{\mathrm{sol}}$ is the solver wall-clock time; and $\psi_t^{\mathrm{time}}=\log(1+\xi_t^{\mathrm{sol}}/\bar\xi^{\mathrm{sol}})$ penalizes slow recovery relative to the reference time, $\bar\xi^{\mathrm{sol}}$. The term $\psi_t^{\mathrm{inf}}$ aggregates load shedding, reserve shortfall, line-flow, ramp, and minimum-time violations returned by the recovery layer. In \eqnref{eq:potential}, $\widehat D_{t+1}^{\mathrm{net}}$ denotes the next-step net-load forecast; $\widehat C^{\mathrm{ED}}(\cdot)$ is a merit-order economic-dispatch surrogate; and $\kappa$ scales the shaping magnitude. The reward is therefore grounded in the physically recovered schedule rather than in a purely learned proxy.

\section{Proposed HSAC-SCUC Method}
\label{sec:method}
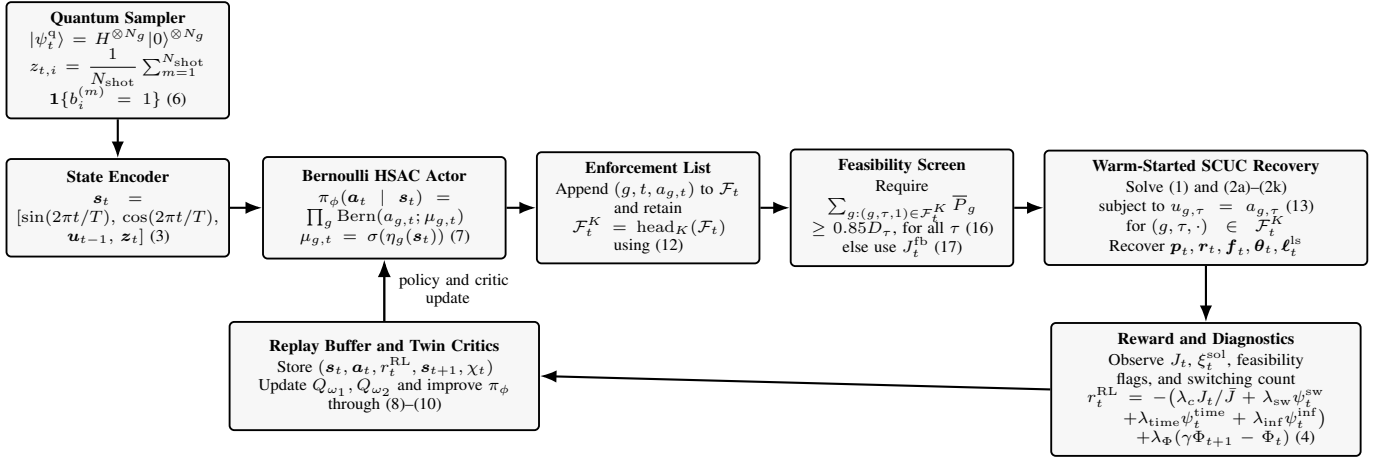
\begin{figure*}[t]
\centering
\resizebox{\textwidth}{!}{\begin{tikzpicture}[
    font=\scriptsize,
    >=Latex,
    node distance=0.42cm and 0.42cm,
    box/.style={draw, rounded corners=2pt, line width=0.8pt, align=center, fill=black!3, inner sep=4pt},
    proc/.style={box, text width=3.05cm, minimum height=1.46cm},
    medproc/.style={box, text width=3.35cm, minimum height=1.52cm},
    wideproc/.style={box, text width=4.55cm, minimum height=1.70cm},
    diagproc/.style={box, text width=4.35cm, minimum height=1.62cm},
    arr/.style={-Latex, line width=0.9pt}
]

\node[proc] (state) {\textbf{State Encoder}\\[1pt]
$\vect{s}_t=[\sin(2\pi t/T),\,\cos(2\pi t/T),$\\
$\vect{u}_{t-1},\,\vect{z}_t]$ \eqnref{eq:state}};

\node[proc, above=0.58cm of state] (qfeat) {\textbf{Quantum Sampler}\\[1pt]
$|\psi_t^{\mathrm{q}}\rangle=H^{\otimes N_g}|0\rangle^{\otimes N_g}$\\
$z_{t,i}=\dfrac{1}{N_{\mathrm{shot}}}\sum_{m=1}^{N_{\mathrm{shot}}}$\\
$\mathbf{1}\{b_i^{(m)}=1\}$ \eqnref{eq:qfeat}};

\node[medproc, right=0.48cm of state] (actor) {\textbf{Bernoulli HSAC Actor}\\[1pt]
$\pi_\phi(\vect{a}_t\mid\vect{s}_t)=\prod_g \operatorname{Bern}(a_{g,t};\mu_{g,t})$\\
$\mu_{g,t}=\sigma\!\left(\eta_g(\vect{s}_t)\right)$ \eqnref{eq:policy}};

\node[proc, right=0.44cm of actor] (fix) {\textbf{Enforcement List}\\[1pt]
Append $(g,t,a_{g,t})$ to $\mathcal{F}_t$\\
and retain $\mathcal{F}_t^K=\operatorname{head}_K(\mathcal{F}_t)$\\
using \eqnref{eq:fixcap}};

\node[proc, right=0.44cm of fix] (screen) {\textbf{Feasibility Screen}\\[1pt]
Require $\sum_{g:(g,\tau,1)\in\mathcal{F}_t^K} \overline P_g$\\
$\ge 0.85D_\tau$, for all $\tau$ \eqnref{eq:filter}\\
else use $J_t^{\mathrm{fb}}$ \eqnref{eq:fallback}};

\node[wideproc, right=0.44cm of screen] (scuc) {\textbf{Warm-Started SCUC Recovery}\\[1pt]
Solve \eqnref{eq:obj} and \eqnref{eq:logic}--\eqnref{eq:security}\\
subject to $u_{g,\tau}=a_{g,\tau}$ \eqnref{eq:fixeq}\\
for $(g,\tau,\cdot)\in\mathcal{F}_t^K$\\
Recover $\vect{p}_t,\vect{r}_t,\vect{f}_t,\vect{\theta}_t,\vect{\ell}_t^{\mathrm{ls}}$};

\node[diagproc, below=0.92cm of actor] (update) {\textbf{Replay Buffer and Twin Critics}\\[1pt]
Store $(\vect{s}_t,\vect{a}_t,r_t^{\mathrm{RL}},\vect{s}_{t+1},\chi_t)$\\
Update $Q_{\omega_1},Q_{\omega_2}$ and improve $\pi_\phi$\\
through \eqnref{eq:target}--\eqnref{eq:pi_loss}};

\node[diagproc, below=0.84cm of scuc] (reward) {\textbf{Reward and Diagnostics}\\[1pt]
Observe $J_t$, $\xi_t^{\mathrm{sol}}$, feasibility flags, and switching count\\
$r_t^{\mathrm{RL}}=-\big(\lambda_cJ_t/\bar J+\lambda_{\mathrm{sw}}\psi_t^{\mathrm{sw}}$\\
$\qquad\ +\lambda_{\mathrm{time}}\psi_t^{\mathrm{time}}+\lambda_{\mathrm{inf}}\psi_t^{\mathrm{inf}}\big)$\\
$\qquad\ +\lambda_{\Phi}(\gamma\Phi_{t+1}-\Phi_t)$ \eqnref{eq:reward}};

\draw[arr] (qfeat) -- (state);
\draw[arr] (state) -- (actor);
\draw[arr] (actor) -- (fix);
\draw[arr] (fix) -- (screen);
\draw[arr] (screen) -- (scuc);
\draw[arr] (scuc) -- (reward);
\draw[arr] (reward.west) -- (update.east);
\draw[arr, line width=1.0pt] (update.north) -- node[right, inner sep=0.5pt, align=center] {\hspace{0.1cm} policy and critic\\update} (actor.south);

\end{tikzpicture}}
\caption{The proposed HSAC-SCUC workflow generates binary commitment proposals, retains at most $K$ chronological commitments for enforcement, rejects obviously undercommitted schedules through a capacity screen, and then recovers dispatch and security variables with the native SCUC model.}
\label{fig:framework}
\end{figure*}

\Figref{fig:framework} summarizes the module-level pipeline, and \Algref{alg:hsac} gives the episode-level training loop. The policy proposes commitments, the capacity screen filters undercommitted subsets, and the native SCUC model returns the recovery-based learning signal.

\subsection{Quantum-Sampled State Augmentation}
The current implementation uses the same number of qubits as generating units, prepared in the shallow reference state $|\psi_t^{\mathrm{q}}\rangle=H^{\otimes N_g}|0\rangle^{\otimes N_g}$ and then measured independently in the computational basis.\footnote{The quantum layer supplies stochastic side information only; feasibility screening and SCUC recovery remain classical.} With $N_{\mathrm{shot}}$ measurement shots, the $i$th auxiliary feature is the empirical marginal
\begin{align}
z_{t,i}=\frac{1}{N_{\mathrm{shot}}}\sum_{m=1}^{N_{\mathrm{shot}}}\mathbf{1}\!\left\{b_i^{(m)}=1\right\}.
\label{eq:qfeat}
\end{align}
This design keeps the quantum channel lightweight while preserving the implemented hybrid interface. Because the present circuit is state-independent, $\vect{z}_t$ should be interpreted as a stochastic auxiliary channel rather than as a quantum optimizer or a trained variational embedding.

\subsection{Bernoulli HSAC for Multi-Binary Commitment}
SAC is adapted to multi-binary UC actions by factorizing the policy across generating units such that
\begin{align}
\pi_\phi(\vect{a}_t\mid\vect{s}_t)=\prod_{g=1}^{N_g}\operatorname{Bern}\!\left(a_{g,t};\mu_{g,t}\right),\quad \mu_{g,t}=\sigma\!\left(\eta_g(\vect{s}_t)\right).
\label{eq:policy}
\end{align}
The actor, therefore, outputs one Bernoulli mean per generating unit. Two critics $Q_{\omega_1}$ and $Q_{\omega_2}$ mitigate overestimation. Using target critics $Q_{\bar\omega_1}$ and $Q_{\bar\omega_2}$, the temporal-difference target is
\begin{align}
y_t &= r_t^{\mathrm{RL}} + \gamma(1-\chi_t)\Big[\min_{j\in\{1,2\}} Q_{\bar\omega_j}(\vect{s}_{t+1},\tilde{\vect{a}}_{t+1}) \nonumber\\
&\hspace{2.0cm}-\alpha\,\mathcal{H}\big(\pi_\phi(\cdot\mid\vect{s}_{t+1})\big)\Big],
\label{eq:target}
\end{align}
where $\alpha$ is the entropy weight, $\chi_t\in\{0,1\}$ is the episode-termination indicator, and $\mathcal{H}(\pi_\phi(\cdot\mid\vect{s}_t))$ is the sum of Bernoulli entropies across units. The critic and actor objectives are
\begin{align}
L_Q &= \mathbb{E}\!\left[\left(Q_{\omega_j}(\vect{s}_t,\vect{a}_t)-y_t\right)^2\right], \label{eq:q_loss}\\
L_\pi &= -\mathbb{E}\!\left[\min_j Q_{\omega_j}(\vect{s}_t,\tilde{\vect{a}}_t)+\alpha\,\mathcal{H}\big(\pi_\phi(\cdot\mid\vect{s}_t)\big)\right].
\label{eq:pi_loss}
\end{align}
In the present implementation, the next-action target uses the Bernoulli mean, and the actor update uses sampled Bernoulli actions. This approximation avoids explicit enumeration of the $2^{N_g}$ action space while retaining entropy regularization \cite{haarnoja2018sac}.

\subsection{Feasibility-Aware Partial Commitment Enforcement}
The key interface between RL and SCUC is the accumulated enforcement list
\begin{align}
\cF_t=\{(g,\tau,a_{g,\tau})\mid g\in\cG,\; \tau\le t\}.
\label{eq:fixlist}
\end{align}
The recovery model does not enforce every historical action. Instead, it retains only the first $K$-tuples,
\begin{align}
\cF_t^K = \firstK(\cF_t),\qquad K=20.
\label{eq:fixcap}
\end{align}
The active enforcement set is imposed through
\begin{align}
u_{g,\tau}=a_{g,\tau},\qquad \forall (g,\tau,a_{g,\tau})\in\cF_t^K.
\label{eq:fixeq}
\end{align}
The resulting recovery problem preserves the original continuous and network-constrained physics, but it reduces the number of free commitment binaries to
\begin{align}
N_{\mathrm{free}}(t)=N_gT-|\cF_t^K|.
\label{eq:nfree}
\end{align}
Equivalently, the fraction of commitment binaries transmitted from the policy to the recovery solver is
\begin{align}
\rho_K(t)=\frac{|\cF_t^K|}{N_gT}\le \frac{K}{N_gT}.
\label{eq:fixratio}
\end{align}
When the cap is active, the unrestricted binary search space is therefore reduced by a factor of $2^{|\cF_t^K|}$ relative to the unfixed commitment block. Before the mixed-integer solution, a capacity-based screen rejects clearly undercommitted schedules via
\begin{align}
\sum_{g:(g,\tau,1)\in \cF_t^K} \overline P_g \ge \beta D_\tau,\qquad \beta=0.85,\;\forall \tau\in\cT.
\label{eq:filter}
\end{align}
If \eqnref{eq:filter} fails, the environment returns a fallback objective
\begin{equation}
J_t^{\mathrm{fb}} = 10^6 + 10^3\sum_{\tau\in\cT}\Bigg[D_\tau - \sum_{g:(g,\tau,1)\in \cF_t^K} \overline P_g\Bigg]_+.
\label{eq:fallback}
\end{equation}
A practically important consequence of the chronological $\firstK$ rule in~\eqnref{eq:fixcap} is its limited temporal coverage under partial commitment enforcement. If each stage contributes $N_g$ new commitment variables, then the number of fully covered periods and the fraction of the next covered period are
\begin{align}
N_{\mathrm{full}}=\left\lfloor\frac{K}{N_g}\right\rfloor,\qquad \sigma_{\mathrm{part}}=\frac{K-N_{\mathrm{full}}N_g}{N_g}.
\label{eq:coverage}
\end{align}
Hence, once $N_g>K$, the recovery model does not receive one fully enforced period. This coverage effect becomes central in the medium-scale results.

\vspace{0.2cm}
\begin{algorithm}[t]
\caption{HSAC-Guided SCUC with Partial Commitment Enforcement}
\label{alg:hsac}
\small
\begin{algorithmic}[1]
\State Initialize actor $\pi_\phi$, critics $Q_{\omega_1},Q_{\omega_2}$, target critics, replay buffer $\mathcal{D}$, and the warm-started SCUC model.
\For{episode $e=1,\ldots,N_{\mathrm{ep}}$}
    \State Reset the environment, obtain $\vect{u}_0$, and set $\cF\leftarrow\emptyset$.
    \For{$t=1,\ldots,T$}
        \State Construct $\vect{s}_t$ using \eqnref{eq:state} and \eqnref{eq:qfeat}.
        \State Sample $\vect{a}_t\sim\pi_\phi(\cdot\mid\vect{s}_t)$ and append $(g,t,a_{g,t})$ to $\cF$ for all $g$.
        \State Keep the active enforcement set $\cF_t^K$ using \eqnref{eq:fixcap}.
        \If{the capacity screen \eqnref{eq:filter} fails}
            \State Assign the fallback objective \eqnref{eq:fallback}.
        \Else
            \State Warm start the SCUC solver, enforce \eqnref{eq:fixeq}, and recover the remaining variables.
        \EndIf
        \State Compute $r_t^{\mathrm{RL}}$ from \eqnref{eq:reward}, store $(\vect{s}_t,\vect{a}_t,r_t^{\mathrm{RL}},\vect{s}_{t+1},\chi_t)$ in $\mathcal{D}$, and update the actor and critics using \eqnref{eq:target}--\eqnref{eq:pi_loss}.
    \EndFor
\EndFor
\end{algorithmic}
\end{algorithm}

\vspace{-.2cm}
\section{Implementation and Experimental Setup}
\vspace{-.1cm}
\label{sec:setup}
\Tabref{tab:impl} summarizes the fixed hyperparameters and solver settings shared by all experiments so that later differences primarily reflect system scale and temporal coverage.
The actor and both critics use two hidden layers with 256 rectified linear unit (ReLU) activations. The replay buffer capacity is $2\times 10^5$, the batch size is 256, the discount factor is $\gamma=0.99$, the target-update rate (SAC target-network soft-update coefficient) is $\upsilon=5\times 10^{-3}$, and the entropy weight is $\alpha=0.05$. The quantum auxiliary channel uses 128 shots per query and was evaluated with IBM Qiskit Aer 0.17.1 using the SamplerV2 primitive. The recovery layer is implemented with UnitCommitment.jl, JuMP, and Gurobi \cite{xavier2024ucjl,lubin2023jump}.
The evaluation uses representative traces for the 14-, 57-, and 118-bus cases, along with budget-sensitivity traces for the same three systems. These instances span the practically relevant transition from $N_g<K$ to $N_g>K$; the 14-bus and 57-bus cases still receive multiperiod commitment enforcement under $K=20$, whereas the 118-bus case can receive only a partial first period. Because the recovery model is kept in the loop, each reported objective is measured after the SCUC recovery layer has enforced dispatch, reserve, and transmission constraints. Recovered-cost statistics are reported separately from capacity-screen rejection events so that penalized early rejections are not conflated with successful SCUC recoveries.

\begin{table}[t]
\caption{Key HSAC-SCUC Implementation Settings}
\label{tab:impl}
\centering
\footnotesize
\setlength{\tabcolsep}{4pt}
\begin{tabularx}{\columnwidth}{@{}lX@{}}
\toprule
\textbf{Component} & \textbf{Setting} \\
\midrule
Actor / critic & Two hidden layers, 256 ReLU units \\
Replay / batch & $2\times 10^5$ transitions / 256 \\
Discount / target update & $\gamma=0.99$, $\upsilon=5\times 10^{-3}$ \\
Entropy weight & $\alpha=0.05$ \\
Reward shaping & Cost, switching, timing, infeasibility, and potential shaping \\
Quantum channel & 128 shots/query; Qiskit Aer 0.17.1; SamplerV2 primitive; shallow Hadamard sampler \\
Recovery solver & \code{UnitCommitment.jl} + \code{JuMP} + \code{Gurobi} \\
Commitment enforcement & Partial commitment enforcement with at most $K=20$ binaries \\
Feasibility screen & Available committed capacity $\ge 0.85\times$ demand \\
Fallback objective & $10^6$ or $10^6 + 10^3\times$deficit \\
\bottomrule
\end{tabularx}
\end{table}

\section{Results and Discussion}
\label{sec:results}
\begin{figure*}[t]
\centering
\hspace{-1.5cm}
\resizebox{0.93\textwidth}{!}{\begin{minipage}[t]{0.290\textwidth}
\centering
\begin{tikzpicture}
\begin{axis}[
    width=0.98\linewidth,
    height=0.426\textwidth,
    scale only axis,
    xmin=0, xmax=1,
    ymin=1.0, ymax=3.6,
    xlabel={Training Progress},
    ylabel={Median / Best},
    ylabel style={font=\footnotesize, xshift=0.2em, xshift=-0.3em},
    xlabel style={font=\footnotesize, yshift=-0.4ex},
    tick label style={font=\footnotesize},
    label style={font=\footnotesize},
    legend style={font=\scriptsize, draw=none, fill=none, at={(0.96,0.73)}, anchor=north east},
    grid=both,
    minor grid style={black!8},
    major grid style={black!15},
    every axis plot/.append style={thick},
    cycle list={blue, red, teal}
]
\addplot+[mark=none] table[x=Progress,y=RollingMedianNorm,col sep=comma] {figures/data/curve_14bus.csv};
\addlegendentry{14-bus}
\addplot+[mark=none, densely dashed] table[x=Progress,y=RollingMedianNorm,col sep=comma] {figures/data/curve_57bus.csv};
\addlegendentry{57-bus}
\addplot+[mark=none, densely dotted] table[x=Progress,y=RollingMedianNorm,col sep=comma] {figures/data/curve_118bus.csv};
\addlegendentry{118-bus}
\end{axis}
\end{tikzpicture}
\end{minipage}
\hspace{0.9cm}
\begin{minipage}[t]{0.255\textwidth}
\centering
\begin{tikzpicture}
\begin{axis}[
    width=0.98\linewidth,
    height=0.486\textwidth,
    scale only axis,
    ymode=log,
    ymin=0.3, ymax=30,
    ylabel={Recovered Cost (\$10$^6$)},
    ylabel style={font=\footnotesize, xshift=-0.3em, yshift=-0.3em},
    xlabel={Case},
    xlabel style={font=\footnotesize, yshift=-0.4ex},
    xtick={1,2,3},
    xticklabels={14,57,118},
    x tick label style={font=\footnotesize},
    tick label style={font=\footnotesize},
    label style={font=\footnotesize},
    grid=both,
    minor grid style={black!8},
    major grid style={black!15},
    boxplot/draw direction=y,
    enlarge x limits=0.18,
    every axis plot/.append style={solid, thick},
    cycle list={blue, red, teal}
]
\addplot+[boxplot prepared={lower whisker=0.3651869549, lower quartile=0.3981803123, median=0.4661221126, upper quartile=0.5583810023, upper whisker=0.7550275385}] coordinates {};
\addplot+[boxplot prepared={lower whisker=1.8861925434, lower quartile=2.6151644126, median=3.2997451355, upper quartile=4.7041145480, upper whisker=19.9789617128}] coordinates {};
\addplot+[boxplot prepared={lower whisker=1.0516236341, lower quartile=1.2701408068, median=1.5782362812, upper quartile=1.9969178293, upper whisker=2.7860090115}] coordinates {};
\end{axis}
\end{tikzpicture}
\end{minipage}
\hspace{1.1cm}
\begin{minipage}[t]{0.280\textwidth}
\centering
\begin{tikzpicture}
\begin{axis}[
    width=0.98\linewidth,
    height=0.446\textwidth,
    scale only axis,
    symbolic x coords={14,57,118},
    xtick=data,
    ymin=0, ymax=5.5,
    xlabel={Case},
    ylabel={$t_{\mathrm{med}}$ (s)},
    ylabel style={font=\footnotesize, xshift=0.2em, yshift=-0.2em},
    xlabel style={font=\footnotesize, yshift=-0.4ex},
    x tick label style={font=\footnotesize},
    tick label style={font=\footnotesize},
    label style={font=\footnotesize},
    grid=both,
    minor grid style={black!8},
    major grid style={black!15},
    every axis plot/.append style={thick}
]
\addplot+[mark=square*] coordinates {(14,0.6821) (57,2.2415) (118,4.6550)};
\end{axis}
\begin{axis}[
    width=0.98\linewidth,
    height=0.446\textwidth,
    scale only axis,
    symbolic x coords={14,57,118},
    xtick=data,
    axis y line*=right,
    axis x line=none,
    ymin=0, ymax=25,
    ylabel={Screen-Rej. (\%)},
    ylabel style={font=\footnotesize, xshift=-0.45em},
    tick label style={font=\footnotesize},
    label style={font=\footnotesize},
    every axis plot/.append style={thick}
]
\addplot+[mark=o, densely dashed] coordinates {(14,3.32) (57,0.92) (118,20.96)};
\end{axis}
\end{tikzpicture}
\end{minipage}}
\caption{The three panels summarize representative HSAC-SCUC experiments on the 14-, 57-, and 118-bus cases. \textit{Left}: a rolling median with a 2\%-of-budget window is computed from recovered episodes only, i.e., episodes that pass the capacity screen and complete the SCUC recovery solution. \textit{Center}: recovered final-cost distributions are shown through 5/25/50/75/95 boxplots in \(\$10^6\). \textit{Right}: the solid square-marked line reports the median per-episode solution time on the left axis, whereas the dashed circle-marked line reports the capacity-screen rejection rate on the right axis, i.e., the percentage of episodes assigned \(J_t^{\mathrm{fb}}\) because the enforced subset does not satisfy \eqnref{eq:filter}.}
\label{fig:results}
\end{figure*}
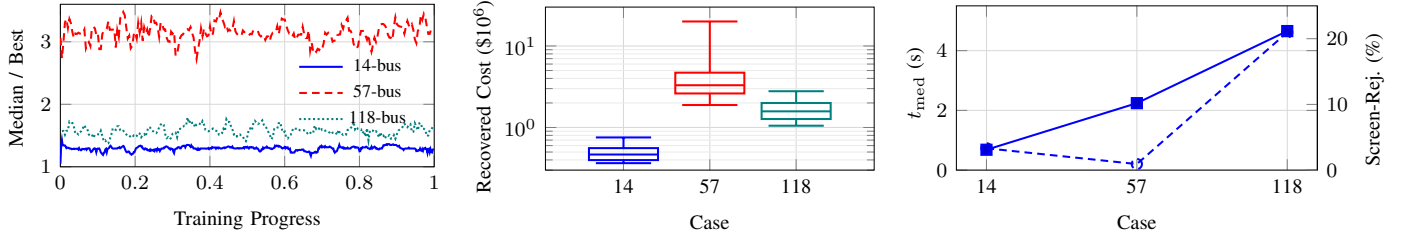

\begin{table}[h]
\caption{Representative HSAC-SCUC Runs and Recovered Costs in \(\$10^6\)}
\label{tab:summary}
\centering
\footnotesize
\setlength{\tabcolsep}{3pt}
\begin{tabular}{lccccccc}
\toprule
\textbf{System} & $N_g$ & \textbf{Ep.} & $J_{\mathrm{best}}^{\mathrm{rec}}$ & $J_{50}^{\mathrm{rec}}$ & $J_{90}^{\mathrm{rec}}$ & $t_{\mathrm{med}}$ & \textbf{Screen-Rej.}\% \\
\midrule
14-bus & 5 & 10{,}000 & 0.361 & 0.466 & 0.673 & 0.68 & 3.32 \\
57-bus & 7 & 5{,}000 & 1.049 & 3.300 & 16.234 & 2.24 & 0.92 \\
118-bus & 54 & 5{,}000 & 1.000 & 1.578 & 2.449 & 4.66 & 20.96 \\
\bottomrule
\end{tabular}
\end{table}

A capacity-screen rejection episode, referred to here as a \textit{fallback episode},\footnote{A fallback episode terminates before the SCUC recovery solution because the committed-capacity screen fails. It is assigned the penalized objective $J_t^{\mathrm{fb}}$ and is excluded from recovered-cost statistics.} is one for which the capacity test in \eqnref{eq:filter} fails, and the environment returns the penalized objective $J_t^{\mathrm{fb}}$ from \eqnref{eq:fallback} without invoking the SCUC recovery solution. \Tabref{tab:summary} and \figref{fig:results} therefore separate two quantities that should not be mixed, namely the quality of the recovered SCUC objective and the frequency with which the enforced subset is rejected before recovery.
The 14-bus case remains the strongest result. Its best recovered objective is \(\$3.61\times 10^5\), its median recovered objective is \(\$4.66\times 10^5\), and its 90th-percentile recovered objective is \(\$6.73\times 10^5\). The left panel of \figref{fig:results} shows that the rolling recovered-cost median settles near $1.30\times$ the best recovered cost, which indicates that the SCUC solver repeatedly receives useful commitment restrictions. The right panel further shows that this behavior is obtained with a subsecond median episode time and a rejection rate of only 3.32\%.
The 57- and 118-bus cases reveal two distinct degradation modes. In the 57-bus case, the capacity-screen rejection rate remains below 1\%, indicating that the proposed commitment subset usually passes the screen. This low rejection rate indicates that the policy has learned commitment patterns that are usually compatible with the fixed startup, shutdown, reserve, and network constraints enforced in the recovery layer. However, the distribution of recovered costs is wide; the median recovered objective is \(\$3.30\times 10^6\), whereas the 90th percentile increases to \(\$1.62\times 10^7\). This indicates that the actor often proposes schedules that are feasible enough to survive screening, but still expensive after the SCUC model reconstructs dispatch, reserves, and line-feasible flows. The larger cost values in the 57-bus case should not be interpreted as a direct statement that it is harder than the 118-bus case.\footnote{Absolute dollar magnitudes are system-specific because the benchmark systems differ in load level, unit fleet, and production-cost coefficients, so cross-system costs should not be compared directly.} In the 118-bus case, by contrast, the rejection rate rises to 20.96\%, but the successful recoveries are comparatively tighter, with a median recovered objective of \(\$1.58\times 10^6\) and a 90th percentile of \(\$2.45\times 10^6\). About 79\% of episodes still pass the capacity screen, indicating that the policy continues to produce recoverable commitment patterns in most accepted episodes. In other words, the 57-bus case is dominated by expensive feasible recoveries, whereas the 118-bus case is dominated by undercoverage of the binary interface itself.
These results clarify both the novelty and the limitations of the proposed approach. Relative to end-to-end RL UC, the policy is not required to satisfy network, reserve, or contingency constraints. Relative to quantum UC formulations based on QAOA, annealing, or QUBO encodings, the method does not attempt to optimize the entire combinatorial block on quantum hardware. The quantum component is only a state-side auxiliary channel. The actual novelty lies in the solver-preserving interface whereby RL proposes commitment binaries, the recovery model enforces only the first $K$ chronological tuples, and the original SCUC solver certifies feasibility within the restricted search space. This positioning is in spirit close to recent solver-compatible commitment-reduction methods \cite{wang2026structure,xiong2025successive}, but the mechanism here is interaction-driven RL rather than supervised masking or linear-programming (LP)-guided restriction.

\begin{figure}[t]
\centering
\resizebox{\columnwidth}{!}{\begin{tikzpicture}
\begin{groupplot}[
    group style={group size=1 by 2, vertical sep=0.72cm},
    width=0.94\columnwidth,
    height=0.42\columnwidth,
    xmode=log,
    log basis x=10,
    xmin=200, xmax=12000,
    grid=both,
    minor grid style={black!8},
    major grid style={black!15},
    tick label style={font=\footnotesize},
    label style={font=\footnotesize},
    every axis plot/.append style={thick},
    cycle list={blue, red, teal}
]
\nextgroupplot[
    ylabel={Median Cost (\$10$^6$)},
    ymin=0.3, ymax=3.7,
    xticklabels=\empty,
    legend style={font=\scriptsize, draw=none, fill=none, at={(0.98,0.90)}, anchor=north east}
]
\addplot+[mark=*] table[x=Episodes,y=MedianCostMillion,col sep=comma] {figures/data/budget_14bus.csv};
\addlegendentry{14-bus}
\addplot+[mark=square*, densely dashed] table[x=Episodes,y=MedianCostMillion,col sep=comma] {figures/data/budget_57bus.csv};
\addlegendentry{57-bus}
\addplot+[mark=triangle*, densely dotted] table[x=Episodes,y=MedianCostMillion,col sep=comma] {figures/data/budget_118bus.csv};
\addlegendentry{118-bus}

\nextgroupplot[
    ylabel={Rejection Rate (\%)},
    xlabel={Episodes},
    xlabel style={yshift=-0.4ex},
    ymin=0, ymax=25
]
\addplot+[mark=*] table[x=Episodes,y=FallbackPct,col sep=comma] {figures/data/budget_14bus.csv};
\addplot+[mark=square*, densely dashed] table[x=Episodes,y=FallbackPct,col sep=comma] {figures/data/budget_57bus.csv};
\addplot+[mark=triangle*, densely dotted] table[x=Episodes,y=FallbackPct,col sep=comma] {figures/data/budget_118bus.csv};
\end{groupplot}
\end{tikzpicture}}
\caption{Training-budget sensitivity for the 14-, 57-, and 118-bus cases. The upper panel reports the median episode objective over all episodes, in \(\$10^6\), whereas the lower panel reports the capacity-screen rejection rate.}
\label{fig:budget}
\end{figure}
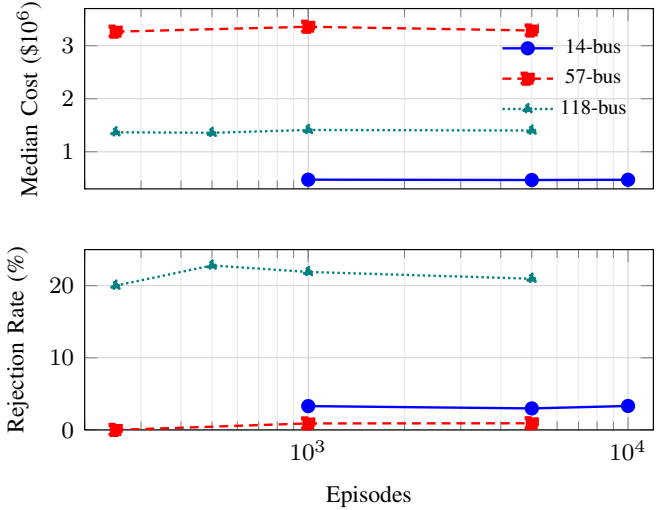

\begin{figure}[h]
\centering
\resizebox{1.02\columnwidth}{!}{\begin{tikzpicture}
\begin{axis}[
    width=0.86\columnwidth,
    height=0.38\columnwidth,
    scale only axis,
    symbolic x coords={14,57,118},
    xtick=data,
    ymode=log,
    ymin=0.2, ymax=10,
    xlabel={Case},
    ylabel={$K/|\mathcal{G}|$},
    xlabel style={font=\small, yshift=-0.4ex},
    ylabel style={font=\small, yshift=-0.2ex},
    tick label style={font=\small},
    label style={font=\small},
    x tick label style={font=\small},
    grid=both,
    minor grid style={black!8},
    major grid style={black!15},
    every axis plot/.append style={thick}
]
\addplot+[mark=diamond*] coordinates {(14,4.000) (57,2.857) (118,0.370)};
\end{axis}
\begin{axis}[
    width=0.86\columnwidth,
    height=0.38\columnwidth,
    scale only axis,
    symbolic x coords={14,57,118},
    xtick=data,
    axis y line*=right,
    axis x line=none,
    ymin=0, ymax=25,
    ylabel={Screen-Rej. (\%)},
    ylabel style={font=\small, xshift=-0.8em},
    tick label style={font=\small},
    label style={font=\small},
    every axis plot/.append style={thick}
]
\addplot+[mark=o, densely dashed] coordinates {(14,3.32) (57,0.92) (118,20.96)};
\end{axis}
\end{tikzpicture}}
\caption{The coverage ratio \(K/|\cG|\) is compared with the capacity-screen rejection rate. The 118-bus case is the first regime in which \(K/|\cG|<1\), i.e., the cap cannot cover a complete period.}
\label{fig:coverage}
\end{figure}
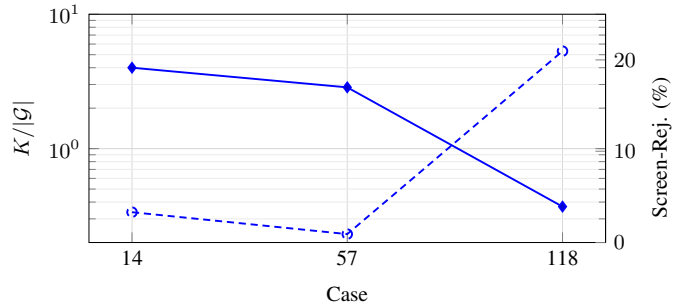

\Figref{fig:budget} shows that longer training budgets do not by themselves remove the dominant error mechanisms. In the 14-bus case, the median episode objective remains in a narrow band around \(\$0.47\times 10^6\) from 1{,}000 to 10{,}000 episodes, and the rejection rate stays close to 3\%. In the 57-bus case, the median episode objective remains near \(\$3.3\times 10^6\) as the budget grows from 250 to 5{,}000 episodes. In the 118-bus case, the median episode objective remains between \(\$1.36\times 10^6\) and \(\$1.40\times 10^6\), whereas the rejection rate stays between 21\% and 23\%. Hence, the current bottleneck is not simply an insufficient number of RL iterations.
A more informative explanation comes from the coverage analysis implied by \eqnref{eq:coverage}. \Tabref{tab:coverage} and \figref{fig:coverage} show that the chronological enforcement rule rapidly loses temporal reach as $N_g$ grows. With $K=20$, the 14-bus case fully covers four periods, and the 57-bus case covers two full periods plus 0.86 of the third period. Once $N_g>K$, however, the recovery model no longer receives even one complete period of commitment enforcement. The 118-bus case is already in this undercovered regime.

\begin{table}[h]
\caption{Coverage Induced by Chronological Partial Commitment Enforcement with \(K=20\)}
\label{tab:coverage}
\centering
\footnotesize
\setlength{\tabcolsep}{4pt}
\begin{tabular}{lcccc}
\toprule
\textbf{System} & $N_g$ & $K/N_g$ & $H_{\mathrm{full}}$ & \textbf{Partial Share} \\
\midrule
14-bus & 5 & 4.000 & 4 & 0.00 \\
57-bus & 7 & 2.857 & 2 & 0.86 \\
118-bus & 54 & 0.370 & 0 & 0.37 \\
\bottomrule
\end{tabular}
\end{table}

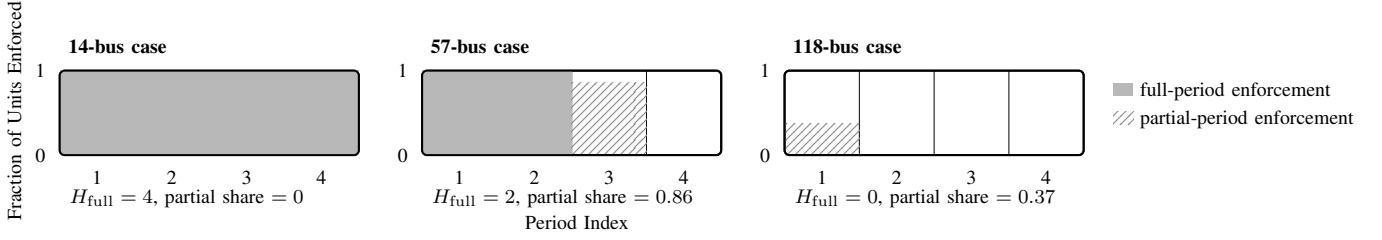
\begin{figure*}[t]
\centering
\resizebox{\textwidth}{!}{\begin{tikzpicture}[font=\footnotesize, x=1cm, y=1cm]
\def\pw{4.25}
\def\ph{1.20}
\tikzset{
  panel/.style={draw, rounded corners=2pt, line width=0.8pt},
  fullfix/.style={fill=black!28, draw=none},
  partfix/.style={pattern=north east lines, pattern color=black!45, draw=none}
}

\newcommand{\drawpanel}[5]{%
  \begin{scope}[shift={(#1,#2)}]
    \draw[panel] (0,0) rectangle (\pw,\ph);
    \foreach \k in {1,2,3} {\draw ({\k*1.0625},0) -- ({\k*1.0625},\ph);} 
    \ifnum#4>0
      \foreach \k in {0,...,\numexpr#4-1\relax} {
        \fill[fullfix] ({\k*1.0625},0) rectangle +(1.0625,\ph);
      }
    \fi
    \pgfmathsetmacro{\xpart}{#4*1.0625}
    \pgfmathsetmacro{\ppart}{#5}
    \ifdim \ppart pt>0pt
      \fill[partfix] (\xpart,0) rectangle +(1.0625,{\ph*\ppart});
    \fi
    \draw[panel] (0,0) rectangle (\pw,\ph);
    \node[font=\footnotesize\bfseries, anchor=south west] at (0,\ph+0.12) {#3};
    \node[anchor=east] at (-0.08,0) {0};
    \node[anchor=east] at (-0.08,\ph) {1};
    \foreach \k/\lab in {0/1,1/2,2/3,3/4} {
      \node[anchor=north] at ({\k*1.0625+0.53125},-0.10) {\lab};
    }
    \node[font=\footnotesize, align=center, anchor=north west] at (0.02,-0.36) {$H_{\mathrm{full}}=#4$, partial share~$=\pgfmathprintnumber[fixed,precision=2]{#5}$};
  \end{scope}}

\drawpanel{0.00}{0.55}{14-bus case}{4}{0.00}
\drawpanel{5.15}{0.55}{57-bus case}{2}{0.86}
\drawpanel{10.30}{0.55}{118-bus case}{0}{0.37}

\node[rotate=90, anchor=south] at (-0.42,1.18) {Fraction of Units Enforced};
\node[anchor=north] at (7.35,-0.18) {Period Index};
\node[anchor=west] at (14.85,1.45) {\tikz{\fill[fullfix] (0,0) rectangle +(0.28,0.18);}~full-period enforcement};
\node[anchor=west] at (14.85,1.05) {\tikz{\fill[partfix] (0,0) rectangle +(0.28,0.18);}~partial-period enforcement};
\end{tikzpicture}}
\caption{Geometry of the chronological partial commitment enforcement for \(K=20\). Solid shading denotes fully enforced periods, and hatched shading denotes the partially enforced period; the transmitted binary information is therefore concentrated on the earliest periods of the horizon.}
\label{fig:geometry}
\end{figure*}

The deterioration in the 118-bus case is not merely a generic RL failure. It is coupled to the combinatorial information that the recovery model actually receives. Under the current chronological enforcement rule, the effective signal passed from the policy to the optimizer becomes very small once the number of units exceeds the cap. The method, therefore, already demonstrates a viable learning-to-optimize interface, but the present subset-selection rule is too coarse for medium-scale SCUC.

\Figref{fig:geometry} makes the temporal asymmetry explicit. Because tuples are appended in period order, the enforced subset always concentrates on the earliest periods of the horizon. This matters for SCUC because minimum up- and downtimes, ramp constraints, reserve trajectories, and contingency-feasible dispatch are all temporally coupled. If only 37\% of period 1 is enforced, as in the 118-bus case, the recovery solver reconstructs nearly the entire commitment trajectory by itself.

\begin{table}[h]
\caption{Coverage and Policy Diagnostics under \(K=20\)}
\label{tab:diag}
\centering
\footnotesize
\setlength{\tabcolsep}{3pt}
\begin{tabularx}{\columnwidth}{@{}lcccX@{}}
\toprule
\textbf{System} & \textbf{On-fraction} & \textbf{Avg. stages} & \textbf{Median shots} & \textbf{Enforcement regime} \\
\midrule
14-bus & 0.501 & 3.98 & 384 & four full periods \\
57-bus & 0.500 & 31.37 & 4{,}608 & two full periods + 0.86 of period 3 \\
118-bus & 0.500 & 36.00 & 4{,}608 & 0.37 of period 1 \\
\bottomrule
\end{tabularx}
\end{table}

\Tabref{tab:diag} reinforces the same conclusion from a different angle. Here, on-fraction denotes the average share of units with $a_{g,t}=1$, average stages denotes the mean number of sequential decision stages completed before termination, and median shots denotes the median number of quantum-measurement shots per episode. It remains essentially 0.5 across all three cases, indicating that the current Bernoulli actor remains highly exploratory. This exploratory behavior preserves feasibility discovery but also slows convergence toward more selective, low-cost subsets as the action space grows. The difference across cases is therefore not driven by a qualitatively different sparsity pattern; instead, it is driven by how much of the sampled binary information the recovery model is allowed to use.

\begin{figure}[t]
\centering
\input{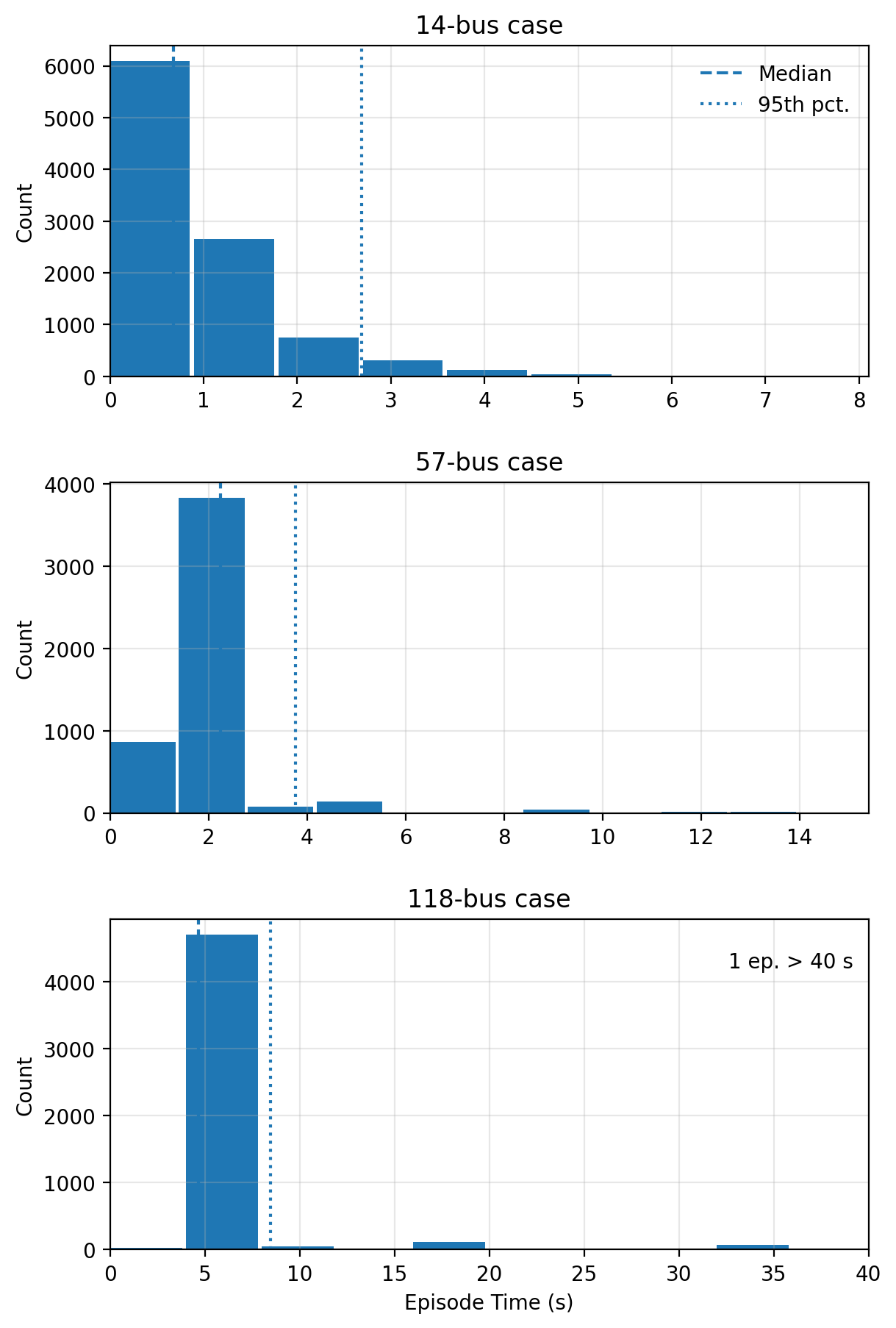}
\caption{Per-episode training-time histograms from the raw summary traces. Dashed and dotted vertical lines mark the median and 95th percentile; one episode above 40 seconds for the 118-bus case is omitted for readability.}
\label{fig:runtime}
\end{figure}

\Figref{fig:runtime} completes the discussion by presenting the per-episode training-time distribution, following the cost, budget, and coverage analyses shown in Figs.~\ref{fig:results}--\ref{fig:geometry}. The histogram for the 14-bus case is broad relative to its median because some episodes recover quickly, whereas others incur additional branch-and-bound work. The histogram for the 57-bus case is sharply centered around 2-3 s, consistent with frequent recovery solutions that are usually feasible but sometimes expensive. The histogram of the 118-bus case is concentrated near 4.5 s to 5 s, but it also exhibits a small secondary cluster in the right tail. This suggests that most recovery solutions behave repeatably once the screen is passed, whereas a small subset incurs substantially higher mixed-integer effort. This interpretation is consistent with the higher rejection rate on the 118-bus case and the tighter recovered-cost distribution in the center panel of \figref{fig:results}. Equation \eqnref{eq:nfree} provides another useful interpretation. For a fixed cap $K$, the recovery layer always removes the same number of binary variables, but the relative impact of that reduction shrinks rapidly as the commitment block grows. Thus, the 14-bus case experiences a meaningful multiperiod reduction of the feasible binary space, the 57-bus case still receives a nontrivial temporal enforcement window, and the 118-bus case experiences only a weak perturbation of a much larger binary polytope. A stronger medium-scale design should therefore adapt $K$ to system size or rank candidate commitment restrictions by operational sensitivity, e.g., ramp-critical, reserve-critical, or congestion-sensitive units.

\section{Conclusion}
\label{sec:conclusion}
This paper presented a three-layer hybrid framework for SCUC in which RL proposes commitment binaries, a quantum-sampled auxiliary channel enriches the state, and a native mixed-integer SCUC model enforces dispatch and security feasibility after partial commitment. The method is scientifically significant because it preserves the original SCUC model, thereby making the learning interface auditable, modular, and compatible with mature, market-oriented optimization software.
The experiments show both promise and clear limits. In the 14-bus case, HSAC-based partial commitment enforcement yields stable low-cost recovery with a sub-second median episode time and a low screen-rejection rate. In the 57-bus case, the recovery model is invoked in almost every episode, yet the recovered-cost distribution remains heavy-tailed, indicating economically weak but feasible commitment subsets. In the 118-bus case, where $N_g>K$, the interface loses complete-period coverage, and the rejection rate rises sharply. The key technical result is therefore the coverage diagnosis: with a fixed cap of $K=20$, the amount of commitment information passed to the solver collapses once the number of units exceeds the cap. Future work will replace the present shallow quantum sampler with higher-fidelity parameterized quantum circuits (PQCs) and pair them with adaptive or confidence-ranked commitment enforcement to improve accuracy and generalization on larger test systems. Additional benchmarking against stronger classical baselines, e.g., learning-based branching, structural masking, successive restriction methods, and hybrid acceleration strategies \cite{dai2025reduction,wang2026structure,xiong2025successive,barrass2025leveraging}, will clarify how much of the remaining limitation is algorithmic and how much is induced by the fixed-cap interface.


\balance
\bibliographystyle{IEEEtran}
\bibliography{References}

@IEEEtranBSTCTL{IEEEexample:BSTcontrol,
  CTLuse_forced_etal       = {yes},
  CTLmax_names_forced_etal = {5},
  CTLnames_show_etal       = {1},
  CTLdash_repeated_names   = {no},
  CTLname_format_string    = {{f.~}{vv~}{ll}{, jj}},
  CTLname_latex_cmd        = {}
}

@article{chen2023scuc,
  author={Chen, Yonghong and Pan, Feng and Qiu, Feng and Xavier, Alinson S and Zheng, Tongxin and Marwali, Muhammad and Knueven, Bernard and Guan, Yongpei and Luh, Peter B. and Wu, Lei and Yan, Bing and Bragin, Mikhail A. and Zhong, Haiwang and Giacomoni, Anthony and Baldick, Ross and Gisin, Boris and Gu, Qun and Philbrick, Russ and Li, Fangxing},
   title   = {Security-Constrained Unit Commitment for Electricity Market: Modeling, Solution Methods, and Future Challenges},
  journal = {IEEE Trans. Power Syst.},
  volume  = {38},
  number  = {5},
  pages   = {4668--4681},
  month   = sep,
  year    = {2023},
  doi     = {10.1109/TPWRS.2022.3213001}
}

@article{yang2021ml,
  author  = {Yafei Yang and Lei Wu},
  title   = {Machine Learning Approaches to the Unit Commitment Problem: Current Trends, Emerging Challenges, and New Strategies},
  journal = {Electr. J.},
  volume  = {34},
  number  = {1},
  note    = {{A}rt. no. 106889},
  month = {Jan.-Feb.},
  year    = {2021},
  doi     = {10.1016/j.tej.2020.106889}
}

@article{xavier2021learning,
  author={Xavier, {\'A}linson S and Qiu, Feng and Ahmed, Shabbir},
  title   = {Learning to Solve Large-Scale Security-Constrained Unit Commitment Problems},
  journal = {INFORMS J. Comput.},
  volume  = {33},
  number  = {2},
  pages   = {739--756},
  year    = {2021},
  doi     = {10.1287/ijoc.2020.0976}
}

@article{pineda2022low,
  author  = {Salvador Pineda and Juan M. Morales},
  title   = {Is Learning for the Unit Commitment Problem a Low-Hanging Fruit?},
  journal = {Electr. Power Syst. Res.},
  volume  = {207},
  note    = {{A}rt. no. 107851},
  year    = {2022},
  doi     = {10.1016/j.epsr.2022.107851}
}

@article{demars2021applying,
  author  = {Patrick de Mars and Aidan O'Sullivan},
  title   = {Applying Reinforcement Learning and Tree Search to the Unit Commitment Problem},
  journal = {Appl. Energy},
  volume  = {302},
  note    = {{A}rt. no. 117519},
  month = nov,
  year    = {2021},
  doi     = {10.1016/j.apenergy.2021.117519}
}

@article{demars2022astar,
  author  = {Patrick de Mars and Aidan O'Sullivan},
  title   = {Reinforcement Learning and {A}* Search for the Unit Commitment Problem},
  journal = {Energy AI},
  volume  = {9},
  note    = {{A}rt. no. 100179},
  month = aug,
  year    = {2022},
  doi     = {10.1016/j.egyai.2022.100179}
}

@article{qin2023ensemble,
  author  = {Jingtao Qin and Yuanqi Gao and Mikhail A. Bragin and Nanpeng Yu},
  title   = {An Optimization Method-Assisted Ensemble Deep Reinforcement Learning Algorithm to Solve Unit Commitment Problems},
  journal = {IEEE Access},
  volume  = {11},
  pages   = {100125--100136},
  year    = {2023},
  doi     = {10.1109/ACCESS.2023.3313998}
}

@article{sayed2024acuc,
  author  = {Ahmed Rabee Sayed and Xian Zhang and Guibin Wang and Yi Wang and Mostafa Shaaban and Mohammad Shahidehpour},
  title   = {Deep Reinforcement Learning-Assisted Convex Programming for {AC} Unit Commitment and Its Variants},
  journal = {IEEE Trans. Power Syst.},
  volume  = {39},
  number  = {4},
  pages   = {5561--5574},
  month = jul,
  year    = {2024},
  doi     = {10.1109/TPWRS.2023.3340674}
}

@article{xu2024wind,
  author  = {Guilei Xu and Zhenjia Lin and Lei Wu and Kwok L. Chan and Junbo Zhang},
  title   = {Deep Reinforcement Learning Based Model-Free Optimization for Unit Commitment Against Wind Power Uncertainty},
  journal = {Int. J. Electr. Power Energy Syst.},
  volume  = {155},
  note    = {{A}rt. no. 109526},
  month = jan,
  year    = {2024},
  doi     = {10.1016/j.ijepes.2023.109526}
}

@article{dai2025reduction,
  author  = {Yuchen Dai and Wei Xu and Minghui Yan and Feng Xue and Jianfeng Zhao},
  title   = {Deep Reinforcement Learning Explanation-Assisted Integer Variable Reduction Method for Security-Constrained Unit Commitment},
  journal = {Eng. Appl. Artif. Intell.},
  volume  = {144},
  note    = {{A}rt. no. 110139},
  month = mar,
  year    = {2025},
  doi     = {10.1016/j.engappai.2025.110139}
}

@article{yan2024lookahead,
  author  = {Jiahao Yan and Yaping Li and Jianguo Yao and Shengchun Yang and Feng Li and Kedong Zhu},
  title   = {Look-Ahead Unit Commitment with Adaptive Horizon Based on Deep Reinforcement Learning},
  journal = {IEEE Trans. Power Syst.},
  volume  = {39},
  number  = {2},
  pages   = {3673--3684},
  month = mar,
  year    = {2024},
  doi     = {10.1109/TPWRS.2023.3286094}
}

@article{liang2024expert,
  author  = {Huijun Liang and Chenhao Lin and Aokang Pang},
  title   = {Expert Knowledge Data-Driven Based Actor-Critic Reinforcement Learning Framework to Solve Computationally Expensive Unit Commitment Problems With Uncertain Wind Energy},
  journal = {Int. J. Electr. Power Energy Syst.},
  volume  = {159},
  note    = {{A}rt. no. 110033},
  month = aug,
  year    = {2024},
  doi     = {10.1016/j.ijepes.2024.110033}
}

@article{venkatesh2025feasibility,
  author  = {Bala Venkatesh and Mohamed Ibrahim Abdelaziz Shekeew and Jessie Ma},
  title   = {Feasibility-Guaranteed Machine Learning Unit Commitment: Fuzzy Optimization Approaches},
  journal = {Appl. Energy},
  volume  = {379},
  note    = {{A}rt. no. 124923},
  month = feb,
  year    = {2025},
  doi     = {10.1016/j.apenergy.2024.124923}
}

@article{lu2026graph,
  author  = {Wentian Lu and Yuexin Zhang and Yihui Zhu and Min Xia and Zheng Han},
  title   = {Graph Reinforcement Learning With Auxiliary Temporal-Graph Convolutional Neural Network for Unit Commitment},
  journal = {Int. J. Electr. Power Energy Syst.},
  volume  = {176},
  note    = {{A}rt. no. 111708},
  month = mar,
  year    = {2026}
}

@ARTICLE{wei2026qrl,
  author={Wei, Xiang and Zhu, Ziqing and Zhu, Linghua and Hu, Ze and Zhang, Xian and Wang, Guibin and Bu, Siqi and Chan, Ka Wing},
  journal={J. Mod. Power Syst. Clean Energy}, 
  title={Quantum Reinforcement Learning Based Two-Stage Unit Commitment with Integration of Virtual Power Plants and Renewable Energy}, 
  year={2026},
  pages={1--12},
  note={early access},
  doi={10.35833/MPCE.2025.000407}}

@article{zheng2024fastquantum,
  author  = {Xiaodong Zheng and Jianhui Wang and Meng Yue},
  title   = {A Fast Quantum Algorithm for Searching the Quasi-Optimal Solutions of Unit Commitment},
  journal = {IEEE Trans. Power Syst.},
  volume  = {39},
  number  = {2},
  month = mar,
  year    = {2024},
  pages={4755-4758},
  doi     = {10.1109/TPWRS.2024.3350382}
}

@article{hong2025qa,
  author       = {Wei Hong and Wangkun Xu and Fei Teng},
  title        = {Qubit-Efficient Quantum Annealing for Stochastic Unit Commitment},
  year         = {2026},
  journal      = {arXiv preprint arXiv:2502.15917v2}
}

@inproceedings{aboumrad2025hybrid,
  author    = {Willie Aboumrad and Phani R. V. Marthi and Suman Debnath and Martin Roetteler and Evgeny Epifanovsky},
  title     = {A New Hybrid Quantum-Classical Algorithm for Solving the Unit Commitment Problem},
  booktitle = {Proc. IEEE Int. Conf. Quantum Comput. Eng. (QCE)},
  pages     = {1905--1915},
  year      = {2025},
  doi       = {10.1109/QCE65121.2025.00208}
}

@article{gao2025dqgbd,
  author  = {Fang Gao and Dejian Huang and Ziwei Zhao and Wei Dai and Mingyu Yang and Qing Gao and Yu Pan},
  title   = {Distributed Quantum Generalized {B}enders Decomposition for Unit Commitment Problems},
  journal = {Quantum Inf. Process.},
  volume  = {24},
  note   = {{A}rt. no. 376},
  year    = {2025},
  doi     = {10.1007/s11128-025-04977-2}
}

@article{hasanzadeh2025d2uc,
  author       = {Milad Hasanzadeh and Amin Kargarian},
  title        = {{D$^2$-UC}: A Distributed-Distributed Quantum-Classical Framework for Unit Commitment},
  year         = {2025},
  journal      = {arXiv preprint arXiv:2511.03104}
}

@article{hasanzadeh2026survey,
  author       = {Milad Hasanzadeh and Amin Kargarian},
  title        = {A Survey on Applications of Quantum Computing for Unit Commitment},
  year         = {2026},
  journal      = {arXiv preprint arXiv:2601.01777}
}

@article{salgado2024hybrid,
  author       = {Bruna Salgado and Andr{\'e} Sequeira and Lu{\'i}s Paulo Santos},
  title        = {A Hybrid Classical-Quantum Approach to Highly Constrained Unit Commitment Problems},
  year         = {2024},
  journal      = {arXiv preprint arXiv:2412.11312}
}

@article{liu2025exact,
  title={Exact quantum algorithm for unit commitment optimization based on partially connected quantum neural networks},
  author={Liu, Jian and Zhou, Xu and Zhou, Zhuojun and Luo, Le},
  journal={Chin. Phys. B},
  volume={34},
  number={10},
  note   = {{A}rt. no. 100303},
  year={2025}
}

@inproceedings{koretsky2021qaoa,
  author    = {Samantha Koretsky and Pranav Gokhale and Jonathan M. Baker and Joshua Viszlai and Honghao Zheng and Niroj Gurung and Ryan Burg and Esa Aleksi Paaso and Amin Khodaei and Rozhin Eskandarpour and Frederic T. Chong},
  title     = {Adapting Quantum Approximation Optimization Algorithm ({QAOA}) for Unit Commitment},
  booktitle = {Proc. IEEE Int. Conf. Quantum Comput. Eng. (QCE)},
  pages     = {181--187},
  year      = {2021},
  doi       = {10.1109/QCE52317.2021.00035}
}

@article{muller2026qa,
  author  = {Sven M{\"u}ller and Marcin Dukalski and Frank Phillipson},
  title   = {Quantum Annealing for Optimizing Unit Scheduling in Renewable Energy Systems: Formulation and Evaluation},
  journal = {IEEE Trans. Power Syst.},
  volume  = {41},
  number  = {2},
  pages   = {836--846},
  month   = mar,
  year    = {2026},
  doi     = {10.1109/TPWRS.2025.3601498}
}

@inproceedings{barrass2025leveraging,
  author    = {Rosemary Barrass and Harsha Nagarajan and Carleton Coffrin},
  title     = {Leveraging Quantum Computing for Accelerated Classical Algorithms in Power Systems Optimization},
editor="Tack, Guido",
title="Leveraging Quantum Computing for Accelerated Classical Algorithms in Power Systems Optimization",
booktitle="Integration of Constraint Programming, Artificial Intelligence, and Operations Research (CPAIOR)",
year="2025",
publisher="Springer Nature Switzerland",
address="Cham",
pages="52--67",
}

@article{feng2023surrogate,
  author  = {Fei Feng and Peng Zhang and Mikhail A. Bragin and Yifan Zhou},
  title   = {Novel Resolution of Unit Commitment Problems Through Quantum Surrogate {L}agrangian Relaxation},
  journal = {IEEE Trans. Power Syst.},
  volume  = {38},
  number  = {3},
  pages   = {2460--2471},
  month = may,
  year    = {2023},
  doi     = {10.1109/TPWRS.2022.3181221}
}

@inproceedings{haarnoja2018sac,
  author    = {Tuomas Haarnoja and Aurick Zhou and Pieter Abbeel and Sergey Levine},
  title     = {Soft Actor-Critic: Off-Policy Maximum Entropy Deep Reinforcement Learning with a Stochastic Actor},
  booktitle = {Proc. 35th Int. Conf. Mach. Learn. (ICML)},
  pages     = {1861--1870},
  year      = {2018}
}

@misc{xavier2024ucjl,
  author       = {Alinson S. Xavier and Aleksandr M. Kazachkov and Ogun Yurdakul and Jun He and Feng Qiu},
  title        = {{UnitCommitment.jl}: A {Julia/JuMP} Optimization Package for Security-Constrained Unit Commitment},
  year         = {2024},
  howpublished = {Zenodo},
  doi          = {10.5281/zenodo.4269874}
}

@article{lubin2023jump,
  author  = {Miles Lubin and Oscar Dowson and Joaquim Dias Garcia and Joey Huchette and Benoit Legat and Juan Pablo Vielma},
  title   = {{JuMP} 1.0: Recent Improvements to a Modeling Language for Mathematical Optimization},
  journal = {Math. Program. Comput.},
  volume  = {15},
  pages   = {581--589},
  year    = {2023},
  doi     = {10.1007/s12532-023-00239-3}
}

@article{wang2026structure,
  author       = {Guangwen Wang and Jiaqi Wu and Yang Weng and Baosen Zhang},
  title        = {Structure-Aware Commitment Reduction for Network-Constrained Unit Commitment with Solver-Preserving Guarantees},
  year         = {2026},
  journal      = {arXiv preprint arXiv:2604.02788}
}

@article{xiong2025successive,
  author       = {Jinxin Xiong and Yanting Huang and Yingxiao Wang and Linxin Yang and Jianghua Wu and Shunbo Lei and Akang Wang},
  title        = {Successive Fixing for Large-Scale Security-Constrained Unit Commitment Using First-Order Methods},
  year         = {2025},
  journal      = {arXiv preprint arXiv:2510.10891}
}

\end{document}